# Large enhancement of spin-flip scattering efficiency at $Y_3Fe_5O_{12}$/Pt interfaces due to vertical confinement

Haripriya Madathil[1], Pranav Pradeep[1], Paul Nöel[2], and Saül Vélez[1,3,*]

[1] Spintronics and Nanodevices Laboratory, Departamento de Física de la Materia Condensada and Condensed Matter Physics Center (IFIMAC), Universidad Autónoma de Madrid, E-28049 Madrid, Spain

[2] Université de Strasbourg, CNRS, IPCMS UMR 7504, Strasbourg F-67034, France

[3] Instituto Nicolás Cabrera, Universidad Autónoma de Madrid, E-28049 Madrid, Spain

*saul.velez@uam.es

Magnons, the quanta of spin angular momentum, can be excited in magnetic insulators by spin-flip scattering processes originated from currents applied to a heavy metal overlayer. The efficiency to generate non-equilibrium magnons across interfaces is parametrized by the spin conductance $g_s$, a phenomenological constant that is considered to be dependent on thermal magnons. Here, we investigate non-linear magnetoresistance phenomena originated in Pt due to current-driven non-equilibrium magnons in $Y_3Fe_5O_{12}$ (YIG). Remarkably, we find that spin-flip scattering processes are dominated by subthermal magnons at room temperature, resulting in a large modulation of $g_s$ with the magnetic field and YIG thickness. Concretely, we find that decreasing the YIG thickness from 100 to 10 nm increases $g_s$ by a factor 30 and observe that the magnetic field exponentially suppresses the magnon generation efficiency. These findings challenge current understanding on $g_s$ and indicate that electrically-driven magnonic effects such as damping compensation and magnon condensation can be largely boosted by device miniaturization.

*Introduction*. Magnetic insulators (MIs) coupled to heavy metals (HMs) are an ideal platform to realize magnonic devices operated by charge currents [1–3]. Diffusive magnons, parametrized by the magnon chemical potential [4], are created in such heterostructures by interfacial spin-flip scattering processes. Most experiments to date have been devoted to investigate the transport characteristics of diffusive magnons by employing non-local lateral structures [5–11], whereas less attention has been given to explore the MI/HM interface spin conductance and emerging phenomena at the metal side [12,13].

Changes in the magnon population result in magnetization variations. Thereby, any magnetoresistance dependent on the magnetization is expected to be modulated when magnons are created or annihilated by spin currents. In a recent work, Nöel et al. show that the creation/annihilation of magnons in Pt/YIG results in a non-linear magnetoresistive response in Pt [13]. However, the underlying physical mechanisms that control the amplitude of this effect remain unclear.

In this work, we investigate non-linear magnetoresistance phenomena stemming from magnon creation/annihilation processes in YIG/Pt of varying YIG thickness and demonstrate that the effect is dominated by subthermal magnons at room temperature. As a result, the interfacial spin-flip scattering efficiency, characterized by the spin conductance $g_s$, is largely modulated by the magnetic field and YIG thickness. Remarkably, we observe that a reduction of YIG thickness from 100 to 10nm results in about two orders of magnitude increase of the magnon modulation efficiency, which we attribute to the vertical confinement of the subthermal magnon modes. These results evidence the potential to realize spin-orbit torque nano-oscillators, two-dimensional magnonic circuits, and explore magnon Bose-Einstein condensate physics at the nanoscale.

*Samples preparation*. Epitaxial YIG films of thicknesses $t = 100, 50, 30, 15,$ and 10nm were grown via sputtering deposition on a (111)-oriented $Ga_5Gd_3O_{12}$ (GGG) substrates as described in Ref. [10]. A 4-nm-thick Pt layer was subsequently sputtered at room temperature in Ar atmosphere at 0.4Pa. Pt Hall bars as the ones shown in Fig. 1a were patterned from the films using optical lithography and Ar plasma etching, leaving the YIG films unetched.

*Experimental setup*. The non-linear magnetoresistive response was explored via harmonic transport measurements in a custom-made magnetotransport setup capable of applying 2T magnetic field and rotate the sample by 360º. An ac current $I$ of frequency $\omega = 10$ Hz is applied and both the longitudinal and transverse responses captured in a high-frequency acquisition card. The harmonic components to the transport (1st and 2nd) are obtained by demodulating the voltage at frequency ω. The first harmonic captures the static magnetoresistive response of the system, whereas the second harmonic reveals non-linear dynamic effects and thermoelectric voltages driven by the current. In particular, in YIG/Pt, the 2nd harmonic signal can be used to characterize current-induced changes in the magnetization arising from the creation and annihilation of magnons [13]. For simplicity, we will focus on the Hall response, although magnon creation/annihilation effects also exhibit a longitudinal counterpart [13]. The mechanism for the emergent non-linear magnetoresistance due to magnons is as follows.

*Non-linear magnetoresistance due to magnon creation/annihilation.* In Pt/YIG the relevant magnetoresistance is the spin Hall magnetoresistance (SMR), which stems from the interaction of the spin currents generated in Pt with the magnetic moments in YIG [14,15]. For the magnetization lying in the plane, the SMR manifests in the Hall response with an angular dependence of the form $\Delta\rho_{xy,\text{SMR}}(\phi) \propto M_x M_y = M_s^2 \cos\phi \sin\phi$, where $\phi$ denotes the angle between the magnetization $\mathbf{M} = (M_x, M_y)$ and the current line. Meanwhile, the magnetization of the YIG layer is modulated by the creation/annihilation of magnons according to $M(I) = M_s + \Delta M(I) \sin(\phi)$, where $\Delta M(I) \propto I$ at moderate currents. The $\sin(\phi)$ describes the angular dependence of the spin-flip efficiency and the sign of the magnetization modulation, which is maxima when **M** is collinear to the spin accumulation in Pt, i.e., for $\phi = 90$ and 270º. Note that the sign of $\Delta M(I) \sin(\phi)$ indicates whether magnons are created (negative) or annihilated (positive), which inverts when the current polarity is inverted or the angle rotated by 180º. To leading order, the modulation of SMR due to magnon creation/annihilation reads as

$$\Delta\rho_{xy,\text{SMR}}(\phi) \propto M_s^2 \cos\phi \sin\phi + 2M_s \Delta M(I) \cos\phi \sin^2\phi$$
$$= R^{1\omega}_{xy,\text{SMR}} \cos\phi \sin\phi + R^{2\omega}_{xy,\text{SMR}} \cos\phi \sin^2\phi, \quad (1)$$

where $R^{1\omega}_{xy,\text{SMR}}$ and $R^{2\omega}_{xy,\text{SMR}}$ indicate the linear and non-linear contributions. The first term is the usual SMR contribution, whereas the second one is associated to the magnon creation/annihilation processes and manifests in the second harmonic due to its linear dependence on the current.

The amplitude of the magnetization variation due to magnon creation/annihilation processes can be evaluated by

$$\frac{\Delta M}{M_s} = R^{2\omega}_{xy,\text{SMR}} / 2 R^{1\omega}_{xy,\text{SMR}}. \quad (2)$$

In YIG/Pt, the second harmonic Hall response $R_{xy}^{2\omega}(\phi)$ includes field-like spin-orbit torque $R_{FL}^{2\omega}$ and spin Seebeck effect $R_{SSE}^{2\omega}$ contributions, whereas the anti-damping torque is negligible due to its weak anomalous Hall effect. Therefore, the angular dependence of $R_{xy}^{2\omega}$ reads as

$$R_{xy}^{2\omega}(\phi) = \left(R_{xy,\text{SMR}}^{2\omega} - R_{xy,\text{FL}}^{2\omega} + R_{xy,\text{SSE}}^{2\omega}\right)\cos\phi + (2R_{xy,\text{FL}}^{2\omega} - R_{xy,\text{SMR}}^{2\omega})\cos^3\phi. \qquad (3)$$

*Modelling $\Delta M/M_s$ from spin-flip scattering at the YIG/Pt interface.* $\Delta M$ is directly proportional to changes in the magnon spectral density $n$. The latter is modified by interfacial spin-flip scattering processes according to [16,17] $\frac{\partial n}{\partial t} = \frac{\partial n}{\partial t_{ref}} + \frac{\partial n}{\partial t_{sc}}$, where $\frac{\partial n}{\partial t_{ref}} = -\frac{n-n_0}{\tau}$ is the magnon relaxation term in relaxation time approximation and $\frac{\partial n}{\partial t_{sc}} = \varepsilon(\Delta\mu_{\uparrow\downarrow} - \Delta\mu_m)n$ the effect of the spin current. $n_0$ is the magnon spectral density at equilibrium, $\tau(\omega)$ the relaxation time, which depends on the frequency $\omega$ of the magnons, $\Delta\mu_{\uparrow\downarrow}$ the current-driven non-equilibrium spin accumulation at the Pt side, $\Delta\mu_m$ the non-equilibrium magnon chemical potential in YIG, and $\varepsilon$ a proportionality factor that depends on the interfacial spin transmission parametrized by the spin conductance $g_s$, following $\varepsilon \propto g_s/t$. The $1/t$ reflects the fact that interfacial changes in the magnon spectral density are averaged across the film thickness. $\Delta\mu_{\uparrow\downarrow}$, to leading order, is proportional to the current and only depends on the charge-to-spin conversion efficiency and the thickness of the metal. Therefore, it can be described by $\Delta\mu_{\uparrow\downarrow} = \beta I$, with $\beta$ a factor that depends on the Pt layer. $\Delta\mu_m$ is also proportional to current with $\Delta\mu_m \lesssim \Delta\mu_{\uparrow\downarrow}$ [4].

The stationary solution for the spectral density is $n(\omega, I) = \frac{n_0}{1-I/I_c}$, where $I_c = 1/\varepsilon\beta(1-\kappa)\tau$ with $\kappa = \Delta\mu_m/\Delta\mu_{\uparrow\downarrow}$. Note that $I_c$ denotes the critical current for damping compensation due to spin pumping into the magnetic layer, where $n$ and thus $\Delta M$, diverge. The variation of the magnon spectral density $\Delta n$ by an alternating current is given by $2\Delta n = n(\omega, I) - n(\omega, -I) = 2n_0 \frac{I/I_c}{1-(I/I_c)^2}$. Therefore $\Delta M/M_s$ reads as:

$$\frac{\Delta M}{M_s} \propto n_0 \frac{\frac{I}{I_c}}{1-\left(\frac{I}{I_c}\right)^2}. \qquad (4)$$

*Harmonic Hall measurements in YIG/Pt.* Magnon creation/annihilation effects, inferred from $R_{xy}^{2\omega}(\phi)$ measurements, are strongly dependent on the magnetic field and YIG thickness. Figs. 1b,c show that at large fields both $R_{xy,\text{FL}}^{2\omega}$ and $R_{xy,\text{SMR}}^{2\omega}$ become negligibly small, and only the spin thermoelectric contribution $R_{xy,\text{SSE}}^{2\omega}$ prevails (black dots). The suppression of $R_{xy,\text{FL}}^{2\omega}$ with field agrees with the $1/B$ dependence of the torques [18] (blue triangles in Fig. 2a), but the suppression of $R_{xy,\text{SMR}}^{2\omega}$ (Fig. 1b) is surprising, as no modulation is expected for thermal magnons. That is because the Zeeman energy $\mu_0 HM$ for a magnetic field $\mu_0 H = 400$mT is orders of magnitude smaller than the thermal energy $k_B T$ at room temperature, indicating that $R^{2\omega}_{xy,\text{SMR}}$ is dominated by deep subthermal magnons, i.e., magnons in the GHz regime with an associated temperature $T^{\text{eff}}$ in the Kelvin regime.

In thin films and for small magnetic fields, $R^{2\omega}_{xy,\text{SMR}}$ is pronounced and dominates $R^{2\omega}_{xy}(\phi)$ (Figs. 1b and 2a). In contrast, in 100-nm-thick YIG, $R^{2\omega}_{xy,\text{SMR}}$ is negligibly small even for small magnetic fields (Fig. 1c), recovering the bulk-like behaviour for $R^{2\omega}_{xy}$ dominated by $R^{2\omega}_{xy,\text{SSE}}$ and $R^{2\omega}_{xy,\text{FL}}$ in YIG/Pt [19]. The strong suppression of $R^{2\omega}_{xy,\text{SMR}}$ in 100nm YIG (~2 orders of magnitude relative to 10nm data) indicate that the efficiency to modulate non-equilibrium magnons by spin currents is strongly influenced by additional effects beyond changes in the magnetic volume (Eqs. 2 and 4). We will address this question in detail below.

An important step to analyse $\frac{\Delta M}{M_s}$ is to unambiguously identify $R^{2\omega}_{xy,\text{SMR}}$ from $R^{2\omega}_{xy}(\phi)$. This can be done by performing $R^{2\omega}_{xy}(\phi)$ measurements at different magnetic fields because each contribution has a different field dependence. In thin films, $R^{2\omega}_{xy,\text{SSE}}$ can be considered constant when the magnetization of the film is saturated [20]. In our analysis we considered a small variation in $R^{2\omega}_{xy,\text{SSE}}(B)$ according to $R^{1\omega}_{xy,\text{SMR}}(B)$, which accounts for the field dependence of $M$ [21], and $R^{2\omega}_{xy,\text{FL}}$ to follow a $1/B$ dependence [18]. A field dependent SSE is further confirmed by independent measurements [21] and found $R^{2\omega}_{xy,\text{FL}}$ to be dominated by the current-induced Oersted field as expected in YIG/Pt [19]. In 10nm YIG, we obtained $B^{\text{FL}} = 0.22 \pm 0.03$mT for $I = 4$mA (blue line Fig. 2a), which is consistent with the Oersted field $B^{\text{Oe}} = \frac{I\mu_0}{2w} = 0.25$mT. Further, we found that $R^{2\omega}_{xy,\text{SMR}}$ follows a power law with the magnetic field $B^{-\gamma}$ (purple line in Fig. 2a, and Fig. 2b), with $\gamma$ ranging between 0.70 at 1mA and 0.83 at 4mA. The distinct field dependence of $R^{2\omega}_{xy,\text{SMR}}$ and $R^{2\omega}_{xy,\text{FL}}$ makes possible to evaluate the magnon contribution to $R^{2\omega}_{xy}$. Furthermore, the non-linear SMR can also be evaluated from $R^{2\omega}_{xx}(\phi)$ measurements. Both analyses produced consistent results [21].

The variation of the magnetization $\Delta M/M_s$ in thin films exhibits a non-linear behaviour with the current and depends on the magnetic field (Fig. 3). The non-linearity is consistent with the current-dependence presented in Eq. (4) (solid lines Fig. 3a). Interestingly, the data shows that in 10-nm-thick-films, damping compensation can be reached even in continuous YIG films [22]. This is possible because the magnon diffusion length $\lambda_m$, which ranges from ~1.3μm to ~300nm in our films [10], is much smaller than the channel width $w$, maximizing the changes in the magnon population underneath the current line. This makes the YIG thickness the only relevant dimensional parameter of the system.

The critical current for damping compensation increases as the magnetic field increases (Fig. 3b). $I_c$ can be accurately estimated for magnetic fields up to ~40mT, following a linear increase with field. At larger fields, however, $\frac{\Delta M}{M_s}(I)$ becomes flatter (Fig. 3a), making the estimate of $I_c$ less reliable (open dots Fig. 3b). The strong field dependence of $I_c$ further supports the idea that the magnons contributing to $\Delta M$ must be deep subthermal. Indeed, the increase of $I_c$ with field is consistent with the expected increase of Gilbert damping for low frequency magnons [23]. The dominant role of subthermal magnons to long-range diffusive transport and current-induced condensation effects are well known [24–27]. Our results demonstrate that subthermal magnons play a key role in spin-flip scattering processes.

$\Delta M$ follows a well-defined exponential dependence with magnetic field $B^{-\gamma}$, with a value for $\gamma$ that strongly depends on the YIG thickness and current applied (Fig. 4). Notably, the unidirectional spin Hall magnetoresistance originated from spin-flip scattering processes in HM/ferromagnet bilayers also exhibits an exponential dependence with field [28], with an amplitude for $\gamma$ following the same trend with $t$ and $I$ as we observe for $\Delta M(B^{-\gamma})$ (Fig. 4c). The dependence of $\gamma(I, t)$ was attributed to changes in the magnon stiffness $D$, which indeed is expected to decrease when either the thickness or current decreases. $D$ can affect $\Delta M$ through $\tau(B)$, yielding $\Delta M \propto 1/(1 + \xi B)$ for low frequency magnons, with $\xi$ a parameter that depends on $D$ [21]. However, such field dependence produce poor fits to the experimental data in contrast to the empirical $B^{-\gamma}$ dependence [Figs. 4a,b], indicating that $D$ is not the dominant parameter that controls $\Delta M(B)$.

The exponential dependence of $\Delta M$ with $B$ is attributed to the field dependence of the spin-flip scattering efficiency, which is parametrized by $g_s(B)$. At $k = 0$, $g_s$ is connected to the magnetization via $g_s \propto \langle M_\perp^2 \rangle$ [15], i.e., the average of the quadratic magnetic components perpendicular to the magnetization. In essence, this term captures the degree of magnetic disorder (magnon occupation) in the system. For instance, $\langle M_\perp^2 \rangle$, and thus $g_s$, increases with current (i.e., temperature) as observed in experiments (Figs. 4a,c and Refs. [21,25,29]). Interestingly, the magnetic field dependence of $\langle M_\perp^2 \rangle$ can be qualitatively estimated from magnetoresistance measurements as follows. Considering $R_{xy,SMR}^{1\omega}(B) = R_{xy,SMR}^{1\omega,sat} - \Delta R_{xy,SMR}^{1\omega}(B)$, the first term denotes the resistivity for $M$ saturated, whereas the second one captures the degree of disorder with $\Delta R_{xy,SMR}^{1\omega}(B) \propto \langle M_\perp^2 \rangle$. Therefore, the field dependence of $g_s$ can be inferred from $\Delta R_{xy,SMR}^{1\omega}(B)$. Remarkably, the analysis reveals an exponential decay of $\Delta R_{xy,SMR}^{1\omega}(B)$ of the form $B^{-\eta}$ with $\eta$ values exhibiting a consistent trend with thickness and current as observed for $\gamma$ [21]. This suggests that the suppression of $\Delta M$ is associated to the modulation of the spin conductance with field, further highlighting the key role of subthermal magnons in spin-flip scattering processes. The demonstrated modulation of $g_s$ with $B$ is also expected to influence other spin-driven magnon effects, such as the amplitude of the magnon signals in non-local devices due to their $g_s^2$ dependence.

Finally, we analyse the role of the YIG thickness on the amplitude of $\Delta M$. Fig. 5a shows the current dependence of $\Delta M$ at $B$ = 10mT, where a very strong thickness dependence can be inferred. Notably, in contrast to the 10nm sample, $I_c$ cannot be experimentally reached for 15-nm-thick YIG. Moreover, for 30, 50 and 100nm, $\Delta M$ is orders of magnitude smaller than for 10nm YIG and exhibits a nearly linear dependence with $I$, indicating that $\frac{I}{I_c} \ll 1$. To quantitatively compare the magnon creation/annihilation efficiencies among the samples, we use Eq. (4). Accordingly, in the linear regime ($\frac{I}{I_c} \ll 1$), $\frac{\Delta M_1}{\Delta M_2} = \frac{\alpha_2 t_2}{\alpha_1 t_1} S_{1,2}$, where the numbers denote the samples and

$$S_{1,2} = \frac{g_{s,1}(1-\kappa_1)n_{0,1}}{g_{s,2}(1-\kappa_2)n_{0,2}}. \qquad (5)$$

Since $\frac{\alpha_2 t_2}{\alpha_1 t_1} \sim 1$ [21,30], the large differences observed in Fig. 5a are essentially captured by $S_{1,2}$ [Eq. (5); see Fig. 5b taking 100nm data as reference] and the non-linearity of $\Delta M(I)$ as $I \to I_c$ in thin films [Eq. (4)]. In bulk materials with same interface quality, $S_{1,2} \sim 1$, indicating that the observed increasing trend of $S_{1,2}$ upon reducing film thickness [Fig. 5b] arises from vertical confinement.

$\kappa$ and $n_0$ are not expected to exhibit large variations among the samples, indicating that $g_s(t)$ must dominate $\Delta M(t)$ [Fig. 5a]. $\kappa$ is proportional to $\Delta \mu_m$, which is a function of $\frac{t}{\lambda_m}$ [4], a value that is roughly constant for all samples because $\lambda_m \propto \alpha^{-1} \propto t$. $n_0$, on the other hand, is expected to be weakly dependent on thickness and exhibit remarkable changes only in the ultrathin limit due to the suppression of the magnon density of states.

It is worth noting that for subthermal magnons, the ones relevant for $\Delta M$, the number of occupied magnon bands [11] $n = 1 + \text{int}\left(\frac{t}{\pi}\sqrt{\frac{k_B T^{\text{eff}}}{\hbar \gamma_m D}}\right)$ reduces to just a few [21]. $k_B T^{\text{eff}}$ refers to the energy of the relevant subthermal magnons, $\gamma_m$ is the gyromagnetic ratio, and $D = 5 \times 10^{-17}$ Tm$^2$ for YIG [31]. We speculate that the increase of $g_s(t)$ upon reducing thickness (Fig. 5b) is due to the vertical confinement of the subthermal magnon modes, leading to the occupation of only a few or a single magnon band. In fact, the 3D to 2D transition is expected to occur at a length scale $\lambda_{\text{Teff}} = \sqrt{\frac{4\pi\hbar\gamma_m D}{k_B T^{\text{eff}}}}$ for magnons of energy $k_B T^{\text{eff}}$. For $T^{\text{eff}} \sim 10K$, $\lambda_{\text{Teff}}$ is $\sim$10nm, which is consistent with the observed steep increase of $S_{1,2} \sim \frac{g_{s,1}}{g_{s,2}}$ in thin films (Fig. 5b), supporting that $g_s$ is dominated by deep subthermal magnons.

*Conclusions.* We analyse changes in the magnetization driven by spin-flip scattering processes at YIG/Pt interfaces and found that the amplitude of the effects is dramatically enhanced in thin films. The phenomenon is explained via the strong thickness and field dependence of the interfacial spin conductance, which to date has been considered to be a phenomenological constant only dependent on temperature when the thermal energy is much larger than the Zeeman energy [4,11]. We attribute this finding to the unexpected dominant role of deep subthermal magnons in spin-flip scattering processes at room temperature. Our work shows that auto-oscillation and magnon condensation effects driven by the spin Hall effect can be dramatically enhanced by geometrical confinement.

**FIGURES**

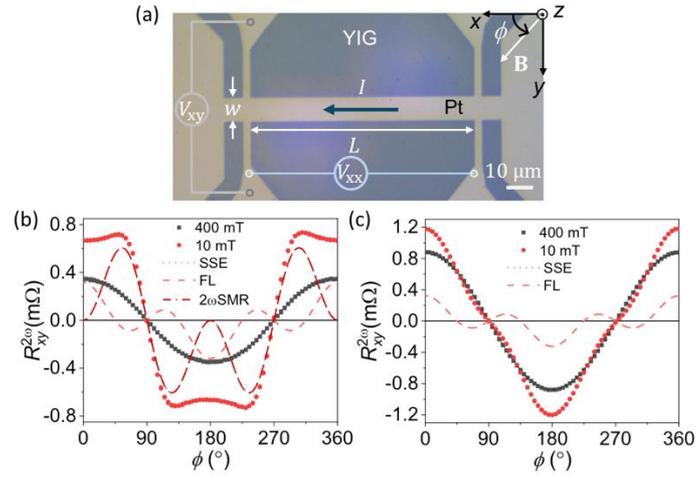

FIG. 1. (a) Optical image of a representative Hall bar device ($w = 10\mu m, L = 100\mu m$) showing the coordinate system, rotation angle and wiring. $R_{xy}^{2\omega}$ is computed from the 2ω voltage response as $V_{xy}^{2\omega}/I$. (b) and (c), $R_{xy}^{2\omega}$ for YIG(10nm)/Pt and YIG(100nm)/Pt, respectively, at 10mT (red dots) and 400mT (black dots) for $I = 4$ mA. The SSE, FL and 2ωSMR curves represent the corresponding contributions obtained from the fit of $R_{xy}^{2\omega}$ using Eq. 3. The SSE curve (black dotted line) overlaps with the $R_{xy}^{2\omega}(400\text{mT})$ experimental data.

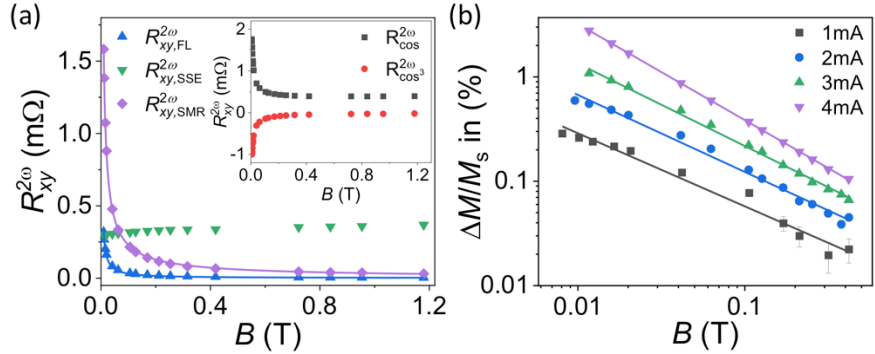

FIG. 2. (a) Field dependence of the different contributions to $R_{xy}^{2\omega}$ in YIG(10nm)/Pt estimated from 2ω angle scans using Eq. (3) (data points). Blue and purple lines show the fit to $B^{-\gamma}$ with $\gamma = 1.00$ and 0.83, respectively. Inset: amplitude of the first (black dots) and second (red dots) terms in Eq. (3). (b) $\Delta M/M_s$ vs $B$ in YIG(10nm)/Pt at different currents computed from $R_{xy,\text{SMR}}^{2\omega}$ using Eq. (2). The solid lines show the fit to $B^{-\gamma}$. Error bars in (a) and (b) are displayed or are smaller than the dot sizes.

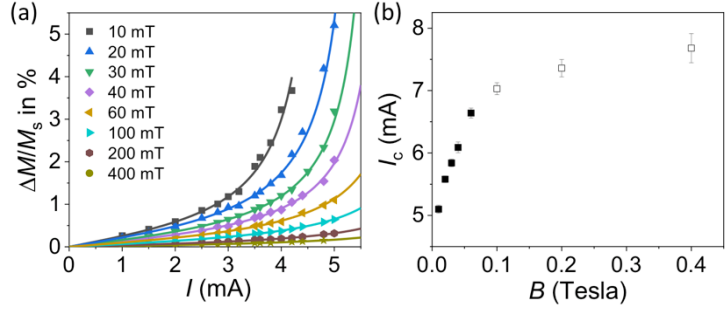

FIG. 3. (a) Current-dependence of $\Delta M/M_s$ in YIG(10nm)/Pt at different magnetic fields. Solid lines show the fits of the data points to Eq. (4). Error bars are smaller than the dots size. (b) Field dependence of $I_c$ extracted from panel (a). Open dots show estimates of $I_c$ for magnetic fields $\geq$ 100mT, where $\Delta M/M_s$ does not exhibit an evident non-linear behaviour. Error bars represent the uncertainty in the estimate of $I_c$ from the fits.

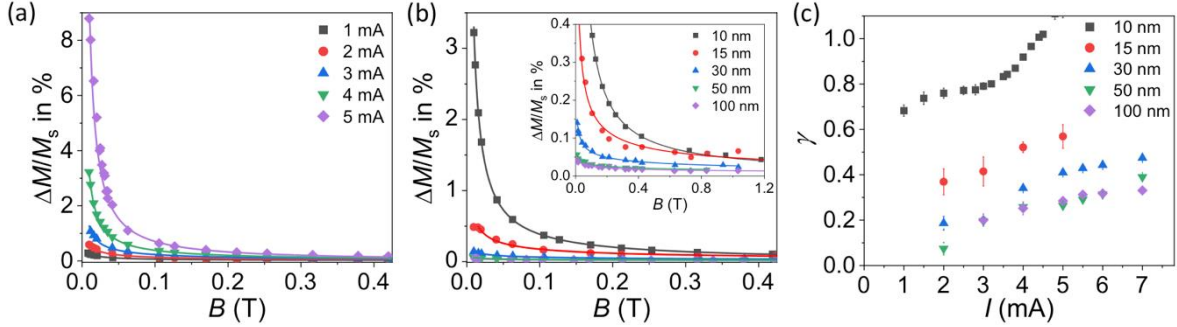

FIG. 4. (a) and (b), Magnetic field dependence of $\Delta M/M_s$ evaluated from $R_{xy}^{2\omega}(\phi)$ (data points) for different currents ($t = 10$nm) and YIG thicknesses ($I = 4$mA), respectively. The solid lines are data fits to $B^{-\gamma}$. Inset in (b): zoom in the small $\Delta M/M_s$ amplitude range. (c) $\gamma$ values extracted from data fits at different currents and for different YIG thicknesses. Error bars are displayed and account for the uncertainty of the fits.

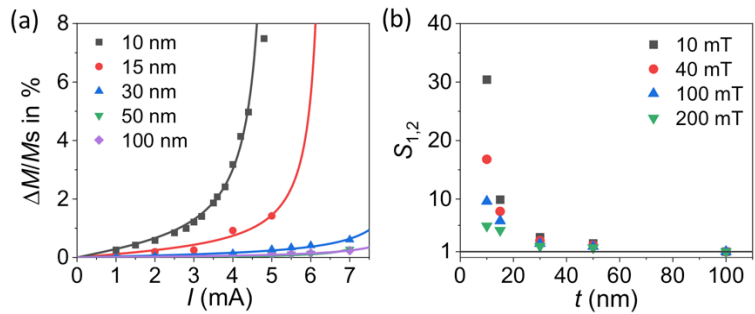

FIG. 5. Current dependence of $\Delta M/M_s$ for different YIG thicknesses. Solid lines present the fits to Eq. (4). (b) Estimates of $S_{12}$ [Eq. (5)] taking 100nm data as reference 2. The non-linearity in $\Delta M/M_s(I)$ due to damping compensation in thin films [Eq. 4] is taken into account to estimate $S_{12}$. An increasing trend in $S_{12}$ upon reducing thickness is observed at all magnetic fields. The smaller $S_{12}(t)$ values at large fields is consistent with the stronger field suppression of $g_s$ as $t$ decreases (Figs. 4b,c).

**ACKNOWLEDGEMENTS**

We acknowledge insightful discussions with A. Kamra. This work was funded by the Spanish Ministry of Science, Innovation and Universities (Grant No. PID2021-122980OA-C53) and by the Comunidad de Madrid through the Atracción de Talento program (Grant No. 2020-T1/IND-20041). H. Madathil acknowledges support from the FPI-UAM PhD scholarship program.